\documentclass[aps]{revtex4}
\usepackage[dvips]{epsfig}
\usepackage{amsmath,amssymb}

\parskip=\medskipamount

\def\be{\begin{equation}}
\def\ee{\end{equation}}
\def\disp{\displaystyle}

\begin{document}

\title{Statistics of randomly branched polymers in a semi-space}
\author{M.V.Tamm$^{1}$, S.K.Nechaev$^{2,3}$, I.Ya.Erukhimovich$^{4}$}
\affiliation{$^{1}$Physics Department, Moscow State University 119992, Moscow,
Russia \\ $^{2}$LPTMS, Universit\'e Paris Sud, 91405 Orsay Cedex, France \\
$^{3}$Landau Institute for Theoretical Physics, 117334, Moscow, Russia \\
$^{4}$A.N.Nesmeyanov Institute of Organoelement Compounds RAS,  Vavilova str. 28,
119991, Moscow, Russia}

\date{September 27, 2004}

\begin{abstract}
We investigate the statistical properties of a randomly branched 3--functional
$N$--link polymer chain without excluded volume, whose one point is fixed at the
distance $d$ from the impenetrable surface in a 3--dimensional space. Exactly
solving the Dyson-type equation for the partition function $Z(N,d)=N^{-\theta}
e^{\gamma N}$ in 3D, we find the "surface" critical exponent $\theta=\frac{5}{2}$,
as well as the density profiles of 3--functional units and of dead ends. Our
approach enables to compute also the pairwise correlation function of a randomly
branched polymer in a 3D semi-space.
\end{abstract}

\maketitle

\section{Introduction}

The rapid development of statistical physics of macromolecules is
indebted, in first turn, to the representation of linear polymers
by Markov chains. The application of a theory of Markov processes
for a description of conformational properties of linear polymer
chains has become invaluable \cite{flory1,gennes,gr_book}. Besides
the linear chains, the polymers of complex geometry, such as, for
example, stars, brushes and branched chains, are of extreme
importance in physical and chemical properties of macromolecular
compounds \cite{flory1,flory2,stock}.

Randomly branched macromolecules are ones of the most interesting
polymer systems with nonlinear structure. The possibility to form
topologically different architectures leads to an extra
contribution to the conformational entropy of samples constituted
by randomly branched chains. There is a deep relation between
physics of randomly branched polymers and "lattice animals"
appeared in percolation \cite{perc1,perc2,luben} and that of the
gelation processes
\cite{csk,rev_kuch,guj,erukh,erukh1,erukh2,sem_rub,pan}. There are
various methods for statistical description of annealed randomly
branched chains. On the one hand, we know successful attempts to
compute various conformational properties of randomly branched
polymers in terms of branched Markov processes
\cite{harris,ikeda,dynkin,kypianou}.

Similarly to the theory of linear polymer chains (see, for
example, \cite{gr_book,RG}), the problem of taking the volume
interactions (i.e. interaction of particles situated topolgically
far from each other) into account plays the central role in the
construction of proper theory for the randomly branched chains.
Concerning this problem, one can imagine two most interesting
limiting cases, corresponding to strong repulsive volume
interactions and negligibly small ones, respectively (we discuss
the applicability of these two limiting cases in the last section
of this paper).

Speaking of the limit of strong repulsive volume interactions, the
special attention should be paid to the application of the
supersymmetry in the description of thermodynamic characteristics
of randomly branched chains. During last two decades since the
seminal works \cite{parisi,shapir,cardy} (see also the short
self-reviews \cite{pcs}) the sypersymmetry becomes a powerful tool
for the investigation of statistics of randomly branched chains.
In particular one has to note the recent contributions
\cite{br_imb,imbrie} which give the impact in understanding the
very origin of the supersymetric dimensional reduction for
randomly branched polymers. Speaking of the main objective of the
present work - the influence of boundary conditions on statistical
properties of randomly branched polymers, we know some recent
successful attempts of calculating the critical exponents of the
randomly branched chains with volume interactions near the
impenetrable wall \cite{debell} and the thorough study of
absorption of a randomly branched polymer \cite{janssen}. The
former of these works is based on rather general lattice
approximation (note that lattice theories always correspond to the
presence of excluded volume), the latter is based on the
supersimmetry technique. In the discussion section we will compare
the results of \cite{debell,janssen} with ours obtained in the
limit of no volume interactions.

In the no interaction (ideal) case the basic question of the
theory concerns the evaluation of the partition function $Z(N)$ of
a non-self-interacting branched structure without loops. Briefly,
the problem is formulated as follows. Take $N$ elementary units
(monomers) such that each monomer has no more than $f_{max}$
branches (we call them $f_{max}$-functional monomers) and compute
all possible ways to arrange them in a {\it single-connected}
loopless cluster. We call such a cluster "a $f_{max}$--branching
random tree". In what follows we shall consider for simplicity
only the case $f_{max}=3$. The problem of calculating the
partition function $Z(N)$ in the free space does not meet any
difficulties and has been solved by various methods (see, for
example, \cite{zimm-stock,daou-joan}). However the influence of
boundary conditions on statistical properties of randomly branched
polymers is far from being as clear as the bulk properties. It
seems to be instructive to compare this situation to the boundary
behavior of linear phantom chains. The boundary effect for linear
random walks can be easily taken into account by using the "image"
principle which enables to represent the partition function of a
linear polymer with, say, Dirichlet boundary conditions in terms
of a linear combination of two shifted bulk partition functions
\cite{spitzer,feller}. {\it A priori} the same "image" method
seems to be inapplicable for randomly branched chains.

In the present work we generalize the technique developed in the
earlier work \cite{erukh,erukh1} for a problem of thermoreversible
gelation to compute a partition function, $Z(N,d)$, and calculate
various thermodynamic characteristics of annealed randomly
branched loopless $N$--link polymer in a 3--dimensional space,
whose one point is fixed at the distance $d$ from the impenetrable
wall. The approach used in our work is based on a diagram
expansion of a partition function of randomly branched polymer in
a semi-space. Summing the diagrams we then construct and solve
directly the Dyson--type equation for the aforementioned partition
function. The developed method enables us also to compute the
density profiles of the monomers having different numbers of
neighbours as well as to derive an expression for the correlation
function, $G(r_1,r_2,N)$, of the branched polymer in a semi-space.

The paper is organized as follows. In the Section II we present the model of
randomly branched polymers and discuss in details the methods we are using. The
brief review of known main results concerning the randomly branched polymers in the
bulk is given in the Section III. The Section IV is devoted to the calculation of a
generating function of branched random walk in a semi-space. That is the central
part of our paper. In the Sections V--VII, using the expression for the generating
function, we compute respectively: the partition function of an $N$--link random
tree, the distribution of branchings in the tree, and the pairwise correlation
function. In the last Section we briefly summarize and discuss the obtained results.

\section{The methods and the model}

In order to make the paper self-consistent, it seems to be instructive to formulate
the general thermodynamic language which has been elaborated for the description of
the thermoreversible gelation process \cite{erukh,erukh1} and is very convenient for
our needs. To begin with, let us consider a system of $N$ identical 3-functional
units capable to form reversible bonds between each other. The partition function
$Z(N)$ of such a system can be written as a product of two terms:
\be
Z(N)=Z_{\rm str}(N) Z_{\rm int}(N)
\label{eq:1}
\ee
where $Z_{\rm str}(N)$ and $Z_{\rm int}(N)$ denote correspondingly the "structural"
and "interactional" parts of the partition function.

The structural contribution to the partition function, i.e. that
due to the formation of clusters of specific structure reads:
\be
Z_{\rm str}(N)=\int \sum_{\{T\}}\frac{1}{r_T}\prod_{\{i,j\}\in T}
\Big[\beta g(r_i,r_j)\Big] dr_i \label{eq:2} \ee where:
\begin{itemize}
\item[--] the product is taken over all pairs of particles forming the manifold
with topology $T$ characterized by the symmetry index $r_T$;
\item[--] the external sum runs over all possible topological structures $T$ of
the system;
\item[--] the factor $\beta$ is the weight of pairing;
\item[--] the function $g(r_i,r_j)=g(r_i-r_j)$ is the probability density to find
two connected particles at the points $r_i$ and $r_j$ correspondingly
\cite{gr_book,lif-gr-kh78,erukh,erukh1}. We assume this function to be the
normalized Gaussian:
\be
g(r)=\left(\frac{3}{2\pi a^2}\right)^{3/2} \exp\left(-\frac{3r^2}{2a^2}\right)
\label{eq:2a}
\ee
where $a$ is a mean-square length of the bond:
\be
a^2=\int r^2 g(r) dr
\label{eq:2b}
\ee
\end{itemize}

The partition function $Z_{\rm int}(N)$ is purely energetic and is due to the
interactions among particles and between particles and an external field:
\be
\begin{array}{l}
\disp Z_{\rm int}(N) =\int  e^{-U(\{r_i\})/T} \prod dr_i \medskip \\
\disp U(\{r_i\})=\sum_i\phi(r_i)+\sum_{i,j}V(r_i,r_j)
\end{array} \label{eq:3}
\ee
where:
\begin{itemize}
\item[--] the potentials $\phi(r_i)$ and $V(r_i,r_j)$ are correspondingly an
external field at the point $r_i$ and a pairwise interaction energy in the system;
\item[--] $T$ is the temperature measured in energetic units.
\end{itemize}
In what follows we neglect the volume interactions between
particles but keep and therefore we set $V(r_i,r_j)=0$.

\begin{figure}[ht]
\epsfig{file=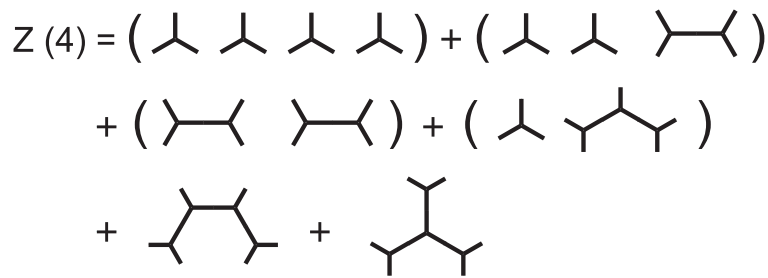,width=10cm} \caption{The diagrammatic representation of the
statistical sum $Z(N)$ for $N=4$.} \label{fig:1}
\end{figure}

In the present article we restrict ourselves to the investigation of randomly
branched polymers only, i.e. we totally neglect the possibility of formation of any
closed loops of bonds in the system. In that case the calculation of the partition
function $Z(N)$ becomes rather transparent and can be described by the simple
diagrammatic technique (see Fig.\ref{fig:1} as an example).

\begin{figure}[ht]
\epsfig{file=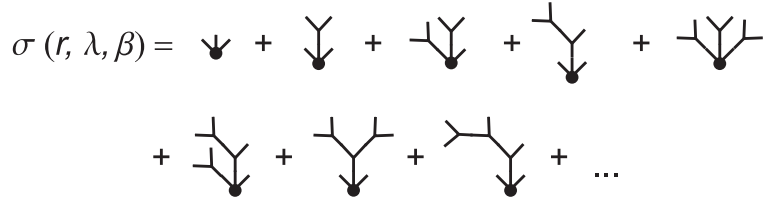,width=10cm} \caption{The series of root diagrams for the
function $\sigma(\lambda,\beta,r)$ up to 4 terms (see Eq.(\ref{eq:5})).}
\label{fig:2}
\end{figure}

Indeed, let us consider a system consisting of $N$ 3-functional monomers bonded in
such a way that no closed loops are present. In general, such a system consists of
many disconnected trees. It is convenient to describe the structure of each tree by
a diagram, where each monomer is represented by a vertex (with an assigned
radius-vector r), the chemical bond between two monomers is represented by a line
connecting the vertices corresponding to the bonded monomers and if a monomer has
less than $3$ bonds with other monomers than each missing bond is represented by a
short line. Next, let us attribute the weights $e^{-\phi(r)/T}$, $\beta
g(r_i,r_j)$ and 1 to the vertices, long lines (connections), and short lines (missing bonds),
respectively. Following the prescription (\ref{eq:2}), the weight of each
diagram is given by a multiplication of all the weights, integration over all space
coordinates and, finally, division by the symmetry index $r_T$. Now, the desired
partition function $Z(N)$ is a sum running over all (connected and disconnected)
weighted diagrams. To proceed further let us introduce the generating function
$\Xi(\lambda)$ of all diagrams (connected and disconnected) as follows:
\begin{eqnarray*}
\Xi(\lambda) & = & 1+\sum_{N=1}^{\infty} Z(N) \lambda^N \medskip \\
Z_{all}(N) & = & \frac{1}{2\pi i}\oint \frac{\Xi(\lambda)}{\lambda^{N+1}}\,d\lambda
\end{eqnarray*}
The function $\Xi(\lambda)$ is just the partition function of the
grand canonical ensemble with a fixed chemical potential $\mu=T\ln
\lambda$. As it follows from the so-called first Mayer theorem
(see, for example, \cite{Ul-Ford}), we can represent
$\Xi(\lambda)$ in the following form: \be
\Xi(\lambda)=\exp{\chi(\lambda,{\phi(r)})} \label{eq:4} \ee where
$\chi(\lambda,{\phi(r)})$ is the generating functional of all the
{\it connected} diagrams (remember that in our approximation all
connected diagrams are just trees). Accordingly, the partition
function of all trees consisting of $N$ monomers is \be Z_{\rm
con}(N)=\frac{1}{2\pi i}\oint
\frac{\chi(\lambda)}{\lambda^{N+1}}\,d\lambda \ee The simplest way
to calculate $\chi(\lambda,{\phi(r)})$ is to define first the
generating function of all the rooted diagrams (i.e. diagrams with
one labelled vertex) $\sigma (\lambda,\beta,r) =-T\delta
\chi(\lambda,{\phi(r)})/\delta \phi(r)$ as it is shown in the
equation below: \be
\begin{array}{l}
\disp \frac {\sigma(\lambda,\beta,r)}{\lambda e^{-\phi(r)/T}} = \frac{1}{6} +
\frac{\lambda\beta}{4}\int g(r,r_1) e^{-\phi(r_1)/T} dr_1 +
\frac{\lambda^2\beta^2}{4} \int g(r,r_1) g(r_1,r_2)
\prod_{i=1}^2 e^{-\phi(r_i)/T} dr_i \\
\disp + \frac{\lambda^2\beta^2}{8} \int g(r,r_1) g(r,r_2) \prod_{i=1}^2
e^{-\phi(r_i)/T} dr_i + \frac{\lambda^3\beta^3}{4} \int g(r,r_1) g(r_1,r_2)
g(r_2,r_3)
\prod_{i=1}^3 e^{-\phi(r_i)/T} dr_i \\
\disp + \frac{\lambda^3\beta^3}{4} \int g(r,r_1) g(r,r_2) g(r_2,r_3) \prod_{i=1}^3
e^{-\phi(r_i)/T} dr_i+ \frac{\lambda^3\beta^3}{48} \int g(r,r_1) g(r,r_2) g(r,r_3)
\prod_{i=1}^3 e^{-\phi(r_i)/T} dr_i \\
\disp + \frac{\lambda^3\beta^3}{16} \int g(r,r_1) g(r_1,r_2) g(r_1,r_3)
\prod_{i=1}^3 e^{-\phi(r_i)/T} dr_i+ \mbox{\{terms of higher orders in $\lambda$ and
$\beta$\}}
\end{array}
\label{eq:5}
\ee
The equation (\ref{eq:5}) can be easily visualized. The corresponding diagrammatic
expansion is displayed in Fig.\ref{fig:2}.

Note that $\lambda^{-1}e^{\phi(r)/T}\sigma(\lambda,\beta,r)$ is a
functional of $\phi(r)$ and a function of $\lambda\beta$ only,
which is due to the fact that the number of bonds in a tree-like
cluster is always one less then the number of connected units:
$N_{bond}=N-1$. Therefore, in what follows we redefine
$$
\epsilon=\lambda\beta
$$
and rewrite $\sigma (\lambda,\beta,r)$ as
\be
\sigma(\lambda,\epsilon,r) = \lambda e^{-\phi(r)/T} \sum_{N=1}^\infty
C_N\epsilon^{N-1} \label{eq:5a}
\ee
It is noteworthy that the coefficient $\lambda e^{-\phi(r)/T} C_N \epsilon^{N-1}$
has a simple physical meaning---it is equal to the partition function of a randomly
branched polymer consisting of $N+1$ links with one link fixed at the point $r$.

Now, it is easy to see that the following relationship holds: \be
\lambda\,d\chi(\lambda,\beta,{\phi(r)})/d\lambda=
\int\sigma(\lambda,r)d^3r \label{z_rho} \ee

Thus, to find the function $\chi(\epsilon)$ we should just
integrate the generating function $\sigma(\lambda,\beta,r)$: \be
\chi(\lambda,\beta)=\beta^{-1} e^{-\phi(r)/T} \sum_{N=1}^\infty
\frac{C_N}{N+1}\beta^N=z^{-1}\int_0^\beta \sigma(\lambda,\tau,r)
d\tau \label{eq:6} \ee

\begin{figure}[ht]
\epsfig{file=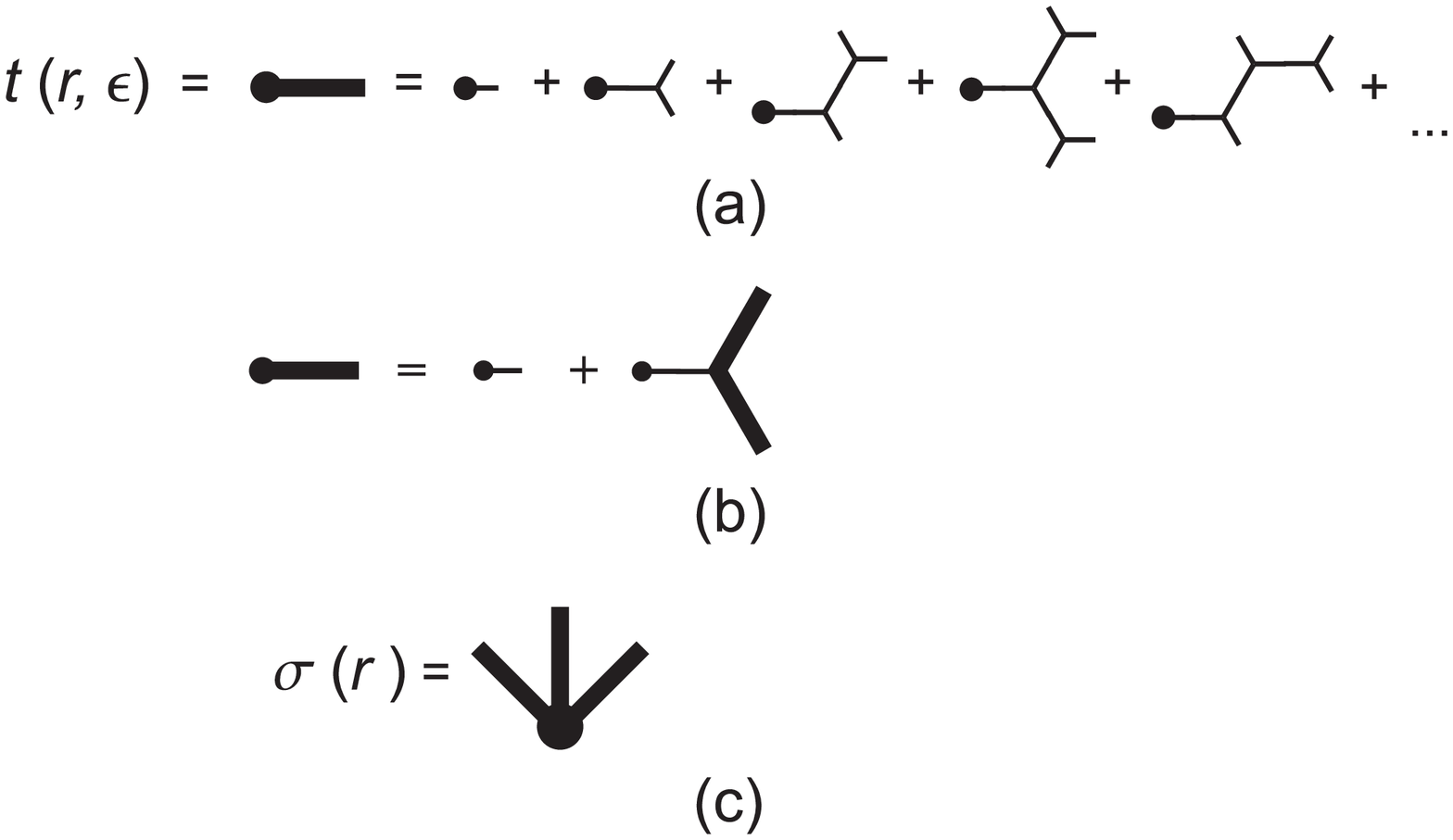,width=9cm} \caption{ Diagrammatic form of: (a)
Eq.(\ref{eq:8}), (b) Eq.(\ref{eq:9}), (c) Eq.(\ref{eq:7}).} \label{fig:3}
\end{figure}

Since the weights of the different branches of any tree are
factorized under assumption of the absence of closed loops and
volume interactions in the system, the generating function $\rho$
can be written in the factorized form: \be
\sigma(\lambda,\epsilon,r) = \frac{\lambda}{6}\,e^{-\phi(r)/T}\,
t^3(\epsilon,r). \label{eq:7} \ee Here $t(\epsilon,r)$ is a
generating function of one branch, which has the following series
expansion: \be
\begin{array}{l}
\disp t(\epsilon,r) = 1 + \frac{\epsilon}{2}\int g(r,r_1) e^{-\phi(r_1)/T} dr_1+
\frac{\epsilon^{2}}{2}\int g(r,r_1) g(r_1,r_2) \prod_{i=1}^2e^{-\phi(r_i)/T} dr_i \\
\disp + \frac{\epsilon^3}{2}\int g(r,r_1) g(r_1,r_2) g(r_2,r_3) \prod_{i=1}^3
e^{-\phi(r_i)/T} dr_i + \frac{\epsilon^3}{8}\int g(r,r_1) g(r_1,r_2) g(r_1,r_3)
\prod_{i=1}^3 e^{-\phi(r_i)/T} dr_i + ...
\end{array}
\label{eq:8} \ee The diagrammatic form of this expansion is shown
in fig. \ref{fig:3}a.

The infinite series (\ref{eq:8}) could be easily evaluated taking into account the
aforementioned statistical independence of different branches. Indeed, it is easy to
see that all the terms in (\ref{eq:8}) are similar to that of the right-hand side of
the diagrammatic equation presented in fig.\ref{fig:3}b. Therefore, the generating
function $t$ satisfies the following exact non-perturbative equation :
\be
t(\epsilon,r) = 1 + \frac{\epsilon}{2}\int g(r,r_1) e^{-\phi(r_1)/T}
t^2(\epsilon,r_1)dr_1. \label{eq:9}
\ee

The equation (\ref{eq:9}) plays the central role in our paper.
Solving it, one gets both "rooted" and "unrooted" generating
functions $\rho$ and $\chi$ and obtains the partition functions
$\Xi$ and $Z$. The method of generating functions enables also to compute
easily the numbers $\rho_{f}$ of the monomers having
exactly $f$ missing bonds (or $f_{max}-f$ bonds with other monomers).
(In what follows we refer to the monomers with $f=0$ and $f=f_{max}-1$ as the junction
monomers and dead ends, respectively.)
 The corresponding generation functions are,
obviously, \be \sigma_f(\lambda,r) = \frac{\lambda\,
e^{-\phi(r_1)/T}}{f!(3-f)!}(t-1)^{3-f}= \lambda\,e^{-\phi(r_1)/T}
\sum_{N=1}^\infty C_N^{(f)}(r)\epsilon ^{N-1} \label{eq:10} \ee
The $N$-th term $C_N^{(f)}(r)$ in the series expansion of $\rho_f$
has a clear physical meaning. It is just a partition function of a
randomly branched $(N+1)$--link polymer with: i) a root fixed at
the given point $r$, and ii) exactly $f$ branches starting from
this root. Comparing this definition with that of $C_N$ (see
equation \ref{eq:5a} and discussion below), we get the probability
$p_f(r)$ for a vertex of a randomly branched $N+1$--link polymer
situated at the point $r$ to be $f$-functional, which equals \be
p_f(r)=\frac{C_N^{(f)}(r)}{C_N(r)} \label{eq:12} \ee

\begin{figure}[ht]
\epsfig{file=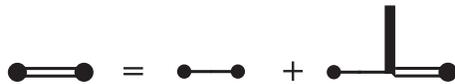,width=6cm} \caption{Visualization of the Dyson equation
Eq.(\ref{eq:12corr}). The correlation function $G$ and the function $g$ are shown
by the double and ordinary solid lines, respectively} \label{fig:4}
\end{figure}

Now, after the generating function $t$ is calculated, one can also calculate readily the
binary correlation function. Indeed, the
correlation function $G(r_1,r_2)$ satisfies the following Dyson equation
\be
G(r_1,r_2)=g(r_1-r_2)+\epsilon \int g(r_1-r_3)t(r_3) e^{-\phi(r_3)/T}G(r_3,r_2)dr_3
\label{eq:12corr}
\ee
which is represented in the diagrammatic form in Fig.\ref{fig:4}.

Note that equation (\ref{eq:12corr}) differs from that suggested
earlier in \cite{daou-joan} where $t(r_3)$ was improperly replaced by $\int
G(r_3,r_4) dr_4$.

\section{The bulk properties of a randomly branched polymer}

In this section we use the formalism introduced above to re-derive the well-known
main characteristics of the randomly branched polymers in the infinite space, which
will be useful to compare with the results in the semi-space to be obtained in the
subsequent sections.

In the infinite homogenous space (i.e. for $\phi(r) \equiv 0$) the equation defining
the generating function of branches $t$ (\ref{eq:9}) has a unique solution which is
invariant with respect to translations and approaches unity when $\epsilon \to 0$.
This solution is easy to find as equation (\ref{eq:9}) becomes purely algebraic due
to the aforementioned translation invariance:
\be
t=\frac{1-\sqrt{1-2\epsilon}}{\epsilon} \label{eq:14}
\ee
where we took into account that $\int g(r) dr = 1$. Note that all other
solutions (in particular that approaching infinity when $\epsilon \to 0$) do
not have any physical meaning and therefore in case of spatially inhomogeneous
external field we will be interested only in solutions of eq.(\ref{eq:9}), which
approach the solution given by Eq. (\ref{eq:14}) when $r \to \infty;\; \phi(r) \to 0$.

The equilibrium densities of junctions and dead ends can be easily
calculated in the way prescribed in the previous section. Indeed,
we have \be \rho=\lambda\,
\frac{t^3}{3!}=\lambda\,\frac{(2-3\epsilon)-
(2-\epsilon)\sqrt{1-2\epsilon}}{3\epsilon^3} \label{eq:rbulk} \ee
for the generating function of all rooted randomly branched trees.
The general expression for randomly branched trees with $f$
missing bonds in the root (compare to (\ref{eq:10})) reads \be
\sigma_f=\lambda\, \frac{(t-1)^{(3-f)}}{f!(3-f)!}
\label{eq:r1bulk} \ee.

Expanding Eqs. (\ref{eq:rbulk}), (\ref{eq:r1bulk}) into series
with respect of $\epsilon$ and substituting the results into eq.(\ref{eq:12}) one
obtains finally the probabilities of junctions and dead ends:
\be
p_0(N)=\frac{(N-2)(N-1)}{2(N+1)(2N+1)};\;\; p_2(N)=\frac{(N+5)}{2(2N+1)}
\label{eq:br_bulk}
\ee
both values approaching $1/4$ as $N$ tends to infinity.

Now, to calculate the correlation function $G(r_1,r_2)$ in the bulk we substitute
Eq. (\ref{eq:14}) into Eq. {\ref{eq:12corr}) and get
\be
G(r)=g(r)+(1-\sqrt{1-2\epsilon})\int g(r')G(r-r')d^3 r' \label{eq:cor_bulk}
\ee
(note that the correlation function in the bulk depends only on the distance between
two roots $r=r_1-r_2$).

There are various methods of solving (\ref{eq:cor_bulk}), we will stick here to the
one which seems to be the most suitable to be generalized in what follows to the case of the semi-space.
As soon as we are interested
mostly in the asymptotic characteristics of the trees when $N$ is large, we can
fully neglect all the short-range (i.e. on the scales of order $r \sim a$)
peculiarities of a correlation function. We therefore replace the first term of the
r.h.s. of (\ref{eq:cor_bulk}) by a Dirac delta-function $\delta(r)$ (see
eqs.(\ref{eq:2a})--(\ref{eq:2b})) and expand the slowly changing function $G(r-r')$
into the series up to the second order in $(r-r')$ (see also \cite{ggsh}). After
calculating the integrals we arrive at the following differential equation:
\be
G(r)=\delta(r)+(1-\xi)\left(G(r)+\frac{a^2}{6}\Delta G(r)\right) \label{eq:cor_sph}
\ee
where we introduced a new variable
$$
\xi=\sqrt{1-2\epsilon}
$$
As the function $G(r)$ depends on $r$ only, we can rewrite the Laplace operator as
$\Delta = r^{-2} \frac{d}{dr} \left(r^2 \frac{d}{dr} \right)$. Denoting
$R=\frac{r\sqrt{6}}{a} \sqrt{\frac{\xi}{1-\xi}}$, we can rewrite the equation
(\ref{eq:cor_sph}) as follows (for $R>0$):
\be
G''(R)+2R^{-1}G'(R)-G(R)=0 \label{eq:cor_sph2}
\ee
Therefore, one gets finally the correlation function in the bulk
\be
G(r)=A\frac{\exp{(-R)}}{R}
\label{eq:cor_sph3}
\ee which allows us to re-derive the well-known result (see
\cite{zimm-stock,daou-joan}) for the gyration radius of the tree:
\be
\left<r^2(\epsilon)\right>=\frac{\int r^2G(r)d^3r}{\int G(r)d^3r} \sim
a^2\left(\xi^{-1}-1\right); \qquad \left<r^2(N)\right> \sim a^2 N^{1/2}
\label{eq:gyr_bulk}
\ee
It is worthwhile also to rewrite the result (\ref{eq:cor_sph3}) in the cylindric
coordinates:
\be
G(Z,\rho)=A\,\frac{\exp{(-\sqrt{Z^2+\rho^2})}}{\sqrt{Z^2+\rho^2}}=
A\int_1^{+\infty}\exp{(-c|Z|)}J_0\left(\rho\sqrt{c^2-1}\right)dc
\label{eq:cor_cyl_bulk}
\ee where $J_0(x)$ is a Bessel function of 0-th order, and the variables $(Z,\,
\rho=\sqrt{x^2+y^2})$ are the usual cylindric coordinates renormalized by the factor
$\frac{\sqrt{6}}{a} \sqrt{\frac{\xi}{(1-\xi)}}$.

\section{Generating function $t$ in a half-space}

Assume now that our branched polymer is displaced in a semi-space $x\ge 0$. The
presence of an impenetrable wall situated at $x=0$ is described by the potential
$\phi(r)\equiv \phi(x,y,z)$, where
\be
\phi(x,y,z) = \left\{\begin{array}{ll} 0 & \mbox{for $x\ge 0$} \\ \infty & \mbox{for
$x<0$} \end{array}\right. \label{eq:15}
\ee
In this case one can rewrite (\ref{eq:9}) as follows
\be
t(x,y,z,\epsilon) = 1 + \frac{\epsilon}{2} \int_{0}^{\infty} dx
\int_{-\infty}^{\infty} dz \int_{-\infty}^{\infty} dy\; g(x,y,z;x',y',z')\, t^2
(x',y',z',\epsilon) \label{eq:16}
\ee
To solve Eq. (\ref{eq:16}) we suggest the following procedure. First of all we represent
(\ref{eq:16}) in the form:
\be \hat{g}^{-1} (t-1) =\frac{\epsilon}{2}\, t^2 \label{eq:17}
\ee
where $\hat{g}^{-1}$ is the inverse operator of $\hat{g}$, the latter being defined
as
$$
\hat{g} f(r) = \int g(r,r') f(r') dr'
$$
Following \cite{lif-gr-kh78} we expand $(t-1)$ in (\ref{eq:17}) in the series up to the second order in $(x-x')$:
\be
t(x',y',z')-1=t(x,y,z)-1+(x-x')\frac{\partial t}{\partial x}+\frac{(x-x')^2}{2}
\frac{\partial^2 t}{\partial x^2}+...\label{eq:17a}
\ee
where we took into account that for obvious physical reasons $t$ depends
neither on $y$ nor on $z$. This expansion is valid for sufficiently smooth function
$t(r)$. Therefore, in the case under consideration this substitution is acceptable
if we are not too close to the surface (the wall). Substituting (\ref{eq:17a}) into
(\ref{eq:17}), we obtain finally
\be
\hat{g}^{-1}  (t-1) \simeq t(x,\epsilon)-1+ \frac{a^2}{6}\frac{\partial^2\,
t(x,\epsilon)}{\partial x^2},\label{eq:17b}
\ee
which results in the following differential equation instead of the integral one:
\be
t(\widetilde{x},\epsilon) - 1 - \frac{\partial^2\,
t(\widetilde{x},\epsilon)}{\partial \widetilde{x}^2}= \frac{\epsilon}{2}\,
t^2(\widetilde{x},\epsilon) \label{eq:19}
\ee
where
$$
\widetilde{x}=\frac{x\sqrt{6}}a
$$
is a reduced distance from the surface.

The differential equation (\ref{eq:19}) can be solved via the substitution $p(t)=
\frac{\partial t}{\partial \widetilde{x}}$. We get after some algebra:
\be
\widetilde{x}(\epsilon,t) = \int_{t_0}^t\frac{dy}{\sqrt{-\frac{2\epsilon}
3y^3+2y^2-4y+\frac{4}{3}\frac{(1-2\epsilon)^{3/2}-1+3\epsilon}{\epsilon^2}}}
\label{eq:19aa}
\ee
where $t_0=t(0,\epsilon) \geq 1$ is a boundary value of $t$ to be specified later
(one could not define this boundary condition {\it a priori}, as the boundary $x=0$
does not belong to the region where the substitution (\ref{eq:17a}) is valid; see
section VI for the discussion of the proper choice of $t_0$) and we have already
used the boundary conditions at $+\infty$:
\be
\left\{\begin{array}{l}
\disp \lim_{x\to +\infty} t = \frac{1-\sqrt{1-2\epsilon}}{\epsilon} \medskip \\
\disp \lim_{x\to +\infty} \frac{dt}{d \widetilde{x}}=0 \end{array}
\right.
\ee
to define the last term in the denominator in (\ref{eq:19aa}).

Performing the substitution
\be
\begin{array}{lll}
\xi & = & \sqrt{1-2\epsilon} \medskip  \\
s & = & 1+(\epsilon y - 1) \xi ^{-1}
\end{array}\label{eq:19b}
\ee
we arrive at the simple integral for the function $\widetilde{x}(\epsilon,t)$:
\be
\widetilde{x}(\epsilon,t) = \sqrt{\frac{3}{2 \xi}}\int\limits_{1+(\epsilon t_0
-1)/\xi} ^{1+(\epsilon t -1)/\xi}\frac{d s}{s \sqrt{3-s}} \label{eq:20}
\ee
which we can easily evaluate:
\be
\widetilde{x}(\xi,t) = \frac{2}{\sqrt{\xi}}\left(\text{arctanh}
\frac{\sqrt{2+4\xi+t(\xi^2-1)}} {\sqrt{6\xi}}-\text{arctanh}
\frac{\sqrt{2+4\xi+t_0(\xi^2-1)}}{\sqrt{6\xi }}\right) \label{eq:21}
\ee
Inverting (\ref{eq:21}) we obtain the desired partition function $t(\widetilde{x})$:
\be
t(\widetilde{x}) = \frac{1}{\epsilon} \left[1-\xi-3\xi\sinh^{-2}
\frac{\widetilde{x} \sqrt{\xi}+ \ln g(\xi,t_0)}2 \right] \label{eq:23}
\ee
where the auxiliary function $g(\xi,t_0)$ reads
\be
g(\xi,t_0) = \frac{\sqrt{2+4\xi + t_0(\xi^2-1)}+\sqrt{6\xi}}
{\sqrt{2+4\xi+t_0(\xi^2-1)}-\sqrt{6\xi}} \label{eq:23a}
\ee

Note, that as $\widetilde{x}$ tends to infinity, the function $\sinh(...)$ in
(\ref{eq:23}) does too and, therefore, the function $t(x)$ approaches its bulk value
(\ref{eq:14}).

\section{Partition function of a randomly branched polymer near the
surface}

To compute the desired partition function $C_N(r)$ of a single randomly branched
polymer, consisting of $N+1$ links, one of which is fixed at a given point $r$, we
should expand the generating function
\be
\rho(\lambda,\epsilon,r) = \frac{\lambda\, t^3(\epsilon,r)}{6} \label{eq:24}
\ee
where $t$ is given by (\ref{eq:23}), in the power series in $\epsilon$. We perform
this expansion in two successive steps. First we expand $t$ in the power series in
$\xi$:
\be
\begin{array}{lll}
t & = & 2(1-3x_0^{-2})  -  \disp  2\left(1-3x_0^{-2}+\frac{x_0^2}5-
\frac{3\sqrt{6}}{5}\,x_0^{-3}\,\frac{5t_0^2+12-10t_0}{(2-t_0)^{5/2}}\right)(1-2\epsilon)
\\  & + & \disp\frac{4}{63}x_0^{-3}\left(x_0^7+2\left(\frac{6}{2-t_0}\right)^{7/2}\right)
(1-2\epsilon)^{3/2}+O\left((1-2\epsilon)^2\right) \label{eq:24a}
\end{array}
\ee
where
\be
x_0=\frac{\widetilde{x}}{2}+\frac{\sqrt{6}}{\sqrt{2-t_0}}
\sqrt{6}\left(\frac{x}{2a}+\frac{1}{\sqrt{2-t_0}}\right) \label{eq:25}
\ee
is a reduced coordinate, and we have used (\ref{eq:19b}) to replace $\xi$ by
$\epsilon$. Thus, the first singular (with respect to $\epsilon$) term in the series
expansion of $t(\epsilon,r)$ is proportional to $(1-2\epsilon)^{3/2}$ and not to
$(1-2\epsilon)^{1/2}$ as in the bulk case. Now, the first singular term in the
series for $\rho(\lambda,\epsilon,r)$ equals to
\be
\rho_{\rm sing}^{(1)}=\frac{8x_0^{-3}}{63}\,z(1-3x_0^{-2})^2
\left(x_0^7+2\left(\frac{6}{2-t_0}\right)^{7/2}\right)(1-2\epsilon)^{3/2}
\label{eq:26}
\ee
Now, taking into account that
$$
(1-2\epsilon)^{3/2}=1-3\epsilon +\frac{3}{2}\epsilon^2+3\sum_{N=3}^{\infty}
\frac{(2N-5)!!}{N!}\epsilon^N
$$
and allowing for the asymptotic behavior of the coefficients in the above sum:
$$
\frac{(2N-5)!!}{N!}=2^N\,N^{-5/2}\left(1+O(N^{-1})\right)
$$
we obtain that the desired partition function $C_n(r)$ for $n\gg 1$ tends to
\be
C_N(x_0) = A(x_0)\, N^{-\theta}\, \epsilon^N \label{eq:28}
\ee
where the function $A(x)$ depends on the microscopic parameters of the model:
\be
A(x_0)=\frac{8}{7}\,x_0^{-3}\left(1-3x_0^{-2}\right)^2
\left(x_0^7+2\left(\frac{6}{2-t_0}\right)^{7/2}\right) \label{eq:28a}
\ee
and the exponent $\theta=\frac{5}{2}$ is universal.

If $x_0$ is large enough (i.e. we are not too close to the surface), one can neglect
all the sub-dominant contributions of $x_0$ and can rewrite
(\ref{eq:28})--(\ref{eq:28a}) in a simpler form
\be
C_N(r) \simeq \frac{2^N}{N^{3/2}}\left(\frac{x^4}{N\,a^4}\right) \label{eq:29}
\ee
Note that the additional (in comparison with the bulk case) factor
$$
\frac{x^4}{N\,a^4}=\left(\frac{\left<x\right>}{a\,N^{1/4}}\right)^{4}
$$
is just the 4-th power of the distance from the surface to the gyration radius of
the randomly branched polymer (\ref{eq:gyr_bulk}). Thus, in contrast to the bulk
behavior, the partition function of the randomly branched polymer near the surface
carries the information about the spatial dimension of the branched polymer.

\section{The probability of branching}

Our next objective is to calculate the probabilities of $f$-functional branchings,
$p_f(r,N)\; (f=1...3)$ in a $N+1$-monomer tree. As we are mostly interested in the
case of very large trees we restrict ourselves here to calculation of the limiting
values $p_f(r)=\lim\limits_{N\rightarrow \infty }p_f(r,N)$. To do that we should
find, according to (\ref{eq:12}), the asymptotic form of the coefficients $C_N(r)$
and $C_N^{(f)}(r)$ in the series expansions of $t^3(r)$ for $C_N(r)$ and of
$(t(r)-1)^3$, $3(t(r)-1)^2$, $3(t(r)-1)$ for $C_N^{(1)}(r)$, $C_N^{(2)}(r)$,
$C_N^{(3)}(r)$ respectively. Thus, we find in particular for the probability of
junctions ($f=3$)
\be
\begin{array}{lll} C_N(r) &=& \disp \frac {4}{7x_0^3}\left(2-\frac
6{x_0^2}\right) ^2\left(x_0^7+2\left( \frac{6}{2-t_0}\right) ^{7/2}\right)
2^N N^{-5/2}\left(1+O\left(N^{-1}\right) \right)  \\
C_N^{(3)}(r) & = & \disp \frac 4{7x_0^3}\left( 1-\frac 6{x_0^2}\right)
^2\left(x_0^7+2\left( \frac 6{2-t_0}\right) ^{7/2}\right) 2^N N^{-5/2}\left(
1+O\left(N^{-1}\right) \right)
\end{array} \label{eq:30}
\ee
and therefore
\be
p_3(r) =\left(\frac{x_0^2-6}{2x_0^2-6}\right)^2
\ee

Similarly, one easily finds for the probability of dead ends ($f=1$):
\be
p_1(r) =\lim_{N\rightarrow \infty}\frac{C_N^{(1)}(r)}{C_N(r)}= \frac{x_0^4}
{\left(2x_0^2-6\right)^2}
\ee

In the Fig. \ref{fig:5} we have plotted the dependencies $p_{1,3}(x_0)$. As $x_0
\rightarrow \infty$ these probabilities approach their bulk values given by
Eq.(\ref{eq:br_bulk}).

\begin{figure}[ht]
\epsfig{file=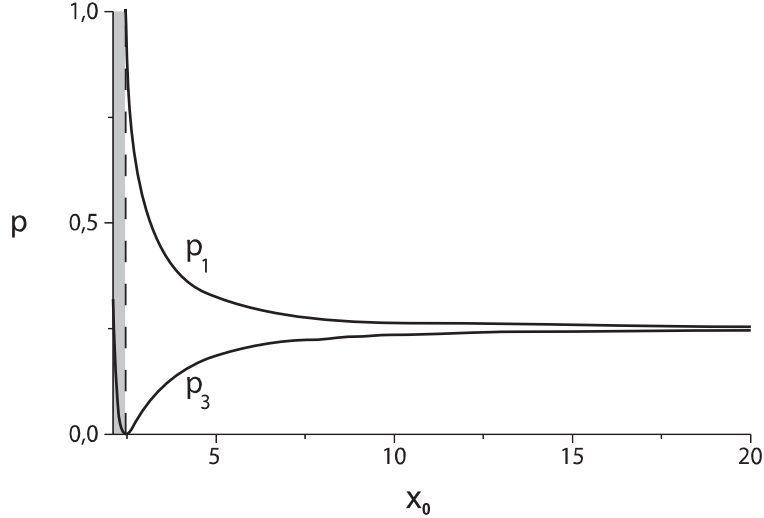,width=10cm} \caption{Profiles of junction points, $p_1$,
and of dead ends, $p_3$. The dashed line corresponds to $x_0=\sqrt{6}$.}
\label{fig:5}
\end{figure}

On the other hand, if one chooses $t_0$, which is the boundary value of the function
$t(x)$ to equal unity ($1\le t(x)\le 2$ for $0\le x <\infty$ (see eq.(\ref{eq:9})),
the value of $x_0$ approaches $\sqrt{6}$ near the wall and the probabilities of dead
ends and junctions tend to unity and zero, respectively. This result seems to be
rather natural: as the wall is impenetrable at $x=0$, there should be mostly dead
ends. Therefore we arrive to the conclusion that the choice of $t_0=1$ is at least
the most natural one thus solving the problem of proper choice of the boundary
conditions at $x=0$ outlined in the discussion after eq. (\ref{eq:19aa}). We
restrict ourselves to this choice in what follows.

\section{The correlation function in the semi-space}

To find the correlation function $G(r_1,r_2)$ of a randomly branched polymer in a
semi-space in 3D we have to solve the equation (\ref{eq:12corr}) where the partition
function $t(r)$ given by eq. (\ref{eq:23}). To do that let us multiply
(\ref{eq:12corr}) by $\epsilon t(r_1)$ and define a new function
$$
\Gamma(r_1,r_2)= \epsilon\, t(r_1) G(r_1,r_2).
$$
After such substitution we get
\be
\Gamma (r_1,r_2)= \epsilon\, t(r_1)\left(g\left(|r_1-r_2|\right)+\int
g\left(|r_1-r_3|\right)\Gamma(r_3,r_2)d^3r_3 \right) \label{eq:35}
\ee
Similarly to what we have done in the previous sections we substitute
$g(|r_1-r_2|)=\delta(r_1-r_2)$ into the first term of the r.h.s. of (\ref{eq:35})
and expand $\Gamma(r_3,r_2)$ up to the second order in $(r_1-r_3)$. Evaluating the
integrals we arrive at the following differential equation
\be
\Gamma(u,v,\rho)=\epsilon\, t(u+v) \left(\Gamma(u,v,\rho) + \Delta_{u,\rho}
\Gamma(u,v,\rho)\right) \label{eq:36}
\ee
where
\be
v=\frac{\sqrt{6} x_2}{a}; \quad u=\frac{\sqrt{6} (x_1-x_2)}{a}; \quad
\rho=\frac{\sqrt{6((y_1-y_2)^2+(z_1-z_2)^2)}}{a} \label{eq:37}
\ee
In what follows we assume $u>0$. Such choice does not lead to any loss of generality
due to the symmetry of the correlation function: $G(r_1,r_2)=G(r_2,r_1)$ and the
case of $u=0$ is to be taken into account via boundary conditions).

Now we seek for the solution of eq.(\ref{eq:36}) in the form
\be
\Gamma (u,v,\rho)=\int f(c,v) \Gamma _1(c,u) \Gamma _2 (c,\rho)dc \label{eq:38}
\ee
where the arbitrary function $f(c,v)$ is to be determined later by the boundary
conditions. The ansatz (\ref{eq:38}) allows us to separate the variables in
(\ref{eq:36}).
We thus obtain a simple equation for $\Gamma_2$
\be
\rho ^{-1} \frac{\partial}{\partial \rho}\left(\rho \frac{\partial
\Gamma_2}{\partial \rho}\right)=c^2\Gamma_2 \label{eq:39}
\ee
the general solution of which is
\be
\Gamma_2(c,\rho)=A\, J_0(c\rho)+B\, Y_0(c\rho) \label{eq:39a}
\ee
where $J_0$ and $Y_0$ are the Bessel functions of the first and second kind. Due to
the boundary condition $\Gamma(c,0)<\infty$, we set $B=0$.

The equation for $\Gamma_1$ is as follows:
\be \frac{\partial^2
\Gamma_1}{\partial u^2}=\left(c^2+\varphi(u,v) \right)
\Gamma_1;\;\;\varphi(u,v)=\frac{1-\epsilon\, t(u+v)}{\epsilon\,
t(u+v)} \label{eq:40} \ee
If we plug the exact expression
(\ref{eq:23}) for $t(u+v)$ into (\ref{eq:40}), the resulting
equation seems to be not solvable. However, if we approximate the genuine
function $\varphi (u,v)$ by its asymptotics in the most interesting regime
$\xi \rightarrow 0$ (i.e. $N \rightarrow \infty$) and $a\xi^{-1/2} \gg x_1 \gg a$, \be \varphi (u,v) \approx \frac{12}{(u+v+p)^2} \label{eq:41} \ee
where
$p=2\sqrt{6}=\lim_{\xi \rightarrow 0}\left(\xi^{-1/2}\ln
g(\xi)\right)$, the resulting differential equation is solvable (by reduction to the Bessel one) precisely.

The corresponding solution is, however, not too accurate. To obtain an improved solution, which approximates
the genuine one in the hole range of variables $u,v$, we are looking for the proper replacement of the genuine function
$\varphi (u,v)$  defined in(\ref{eq:40}) in the form
$$
\widetilde{\varphi}(u,v) = \frac{c_1(v)}{(u+c_2(v))^2}
$$
thus preserving the limiting behavior of $\varphi$ at $u \rightarrow \infty,\,\xi \rightarrow 0$.
To preserve the limiting behavior
of (\ref{eq:40}) at $u \rightarrow 0$ we now need to set
$$
\left\{\begin{array}{l}
\varphi(u=0,v)=\widetilde{\varphi}(u=0,v) \medskip \\
\disp \frac{\partial \varphi(u,v)}{\partial u}\bigg|_{u=0} = \frac{\partial
\widetilde{\varphi}(u,v)}{\partial u}\bigg|_{u=0}
\end{array}\right.
$$
which results into
\be
\widetilde{\varphi}(u) \simeq \frac{12((p+v)^2-12)}{((p+v)(u+p+v)-12)^2}
\label{eq:42}
\ee
At last, in the limit of $u \gg \xi^{-1/2}$  one should
replace $\varphi$ by its bulk value $\varphi=\xi /(1-\xi)$ which does not depend on
$u$ and therefore does not affect the desired solubility of the differential equation (\ref{eq:40}). Thus, finally,
\be
\widetilde{\varphi}(u)=\frac{\xi}{1-\xi}+\frac{12((p+v)^2-12)}{((p+v)(u+p+v)-12)^2}
\label{eq:42a}
\ee
seems to satisfy the desired conditions, both being a good approximation of $\varphi$ in the whole
range of parameters and making the eq. (\ref{eq:40}) solvable. The
comparison of the functions $\widetilde{\varphi}(u,v)$ and $\varphi(u,v)
=\frac{1-\epsilon\, t(u+v)}{\epsilon\, t(u+v)}$ for two different values of the
parameter $v$ ($v=0$ and $v=3$) is shown in the Fig.\ref{fig:6}.

\begin{figure}[ht]
\epsfig{file=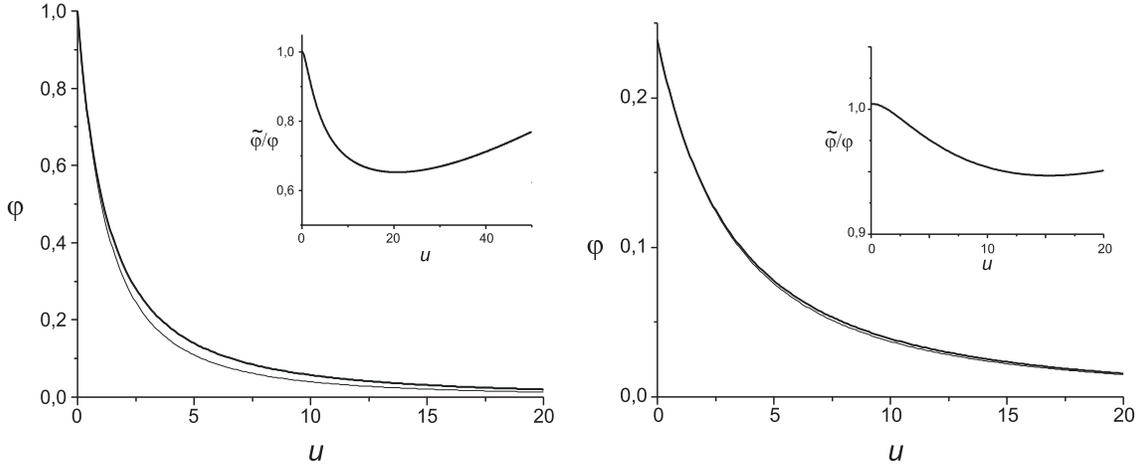,width=15cm} \caption{Plots of the functions
$\widetilde{\varphi}(u,v)$ (thin line) and $\varphi(u,v) =\frac{1-\beta
t(u+v)}{\beta t(u+v)}$ (bold line) for $v=0$ (left) and for $v=3$ (right). The
insertions in both plots show the ratio $\widetilde{\varphi}/\varphi$.}
\label{fig:6}
\end{figure}

As one sees, the approximation of the exact function $\varphi (u,v)$ by the function
$\widetilde{\varphi}(u,v)$ is reasonable for $v=0$ and already very good for $v=3$.
Let us recall that the distance from the wall $v=3$ in the non-renormalized
(initial) coordinates according to (\ref{eq:37}) is $x_2=3a/\sqrt{6}\approx 1.22 a$.

The resulting differential equation
$$
\Gamma_1''=(c^2+\widetilde{\varphi}(u)) \Gamma_1
$$
is solved precisely via substitution $w=u+v+p-\frac{12}{v+p}$.
Taking into account the boundary condition $\Gamma_1 \rightarrow
0$ at infinity, one obtains the final result in the following form
\be \Gamma_1= \sqrt{p w}\, K_{\nu}(c'w)
\label{eq:43}
\ee
where $K_{\nu}(c'v)$ is a modified Bessel function of 2-nd kind of order $\nu$, and
the parameters are as follows:
\be \nu=\frac{\sqrt{49(v+p)^2-576}}{2(v+p)}; \quad c'=\sqrt{c^2+\frac{\xi}{1-\xi}}
\label{eq:43a}
\ee

Substituting (\ref{eq:39a}), (\ref{eq:43}) into (\ref{eq:38}) we arrive at the
desired solution of the equation (\ref{eq:36}):
\be
\Gamma(u,v,\rho)=\int f(c,v)\; J_0(c\rho)\; \sqrt {p w}\, K_{\nu}(c'w)\;dc
\label{eq:44}
\ee
To make this equation comparable with that in the bulk case
(\ref{eq:cor_cyl_bulk}) let us introduce the new variables:
\be
C=\sqrt{\frac{c^2(1-\xi)}{\xi}+1}; \quad
\left\{\begin{array}{c} R \\ U \\ V \\ W \\ P \end{array}\right\}=\sqrt{\frac{\xi}{1-\xi}}
\left\{\begin{array}{c} \rho \\u \\v \\ w \\ p \end{array} \right\}
\ee
After such a substitution the correlation function acquires the form
\be
\Gamma(U,V,R)=\int f(C,V)\; J_0(R\sqrt{C^2-1})\;\sqrt{W} \;K_{\nu}(C W)\; dC
\ee
The unknown function $f(C,V)$ is to be determined by the boundary conditions
\be
\Gamma(U,V,R)\bigg|_{U\gg 1 \; \text{or} \; V \gg 1} \to \epsilon\; t_{\rm bulk}\;
G_{\rm bulk}(\sqrt{U^2+R^2}) \label{eq:45}
\ee
One thus easily finds
\be
f(C,V) \sim \sqrt{C}\exp{\left[C\left(V+P-\frac{12}{V+P}\right)\right]}, \quad  C>1
\label{eq:46}
\ee
where we have omitted all terms which do not depend on $c$ and $v$. Let
us stress that the condition $C>1$  ultimately defines the limits of integration in
(\ref{eq:46}). Thus we arrive finally at the following expression for the
correlation function
\be
G(u,v,\rho)=\int_{1}^{+\infty}\sqrt{C W}e^{C(V+P-12(V+P)^{-1})}K_{\nu}(C W)
J_0\left(R\sqrt{C^2-1}\right)d C \label{eq:47}
\ee

\begin{figure}[ht]
\epsfig{file=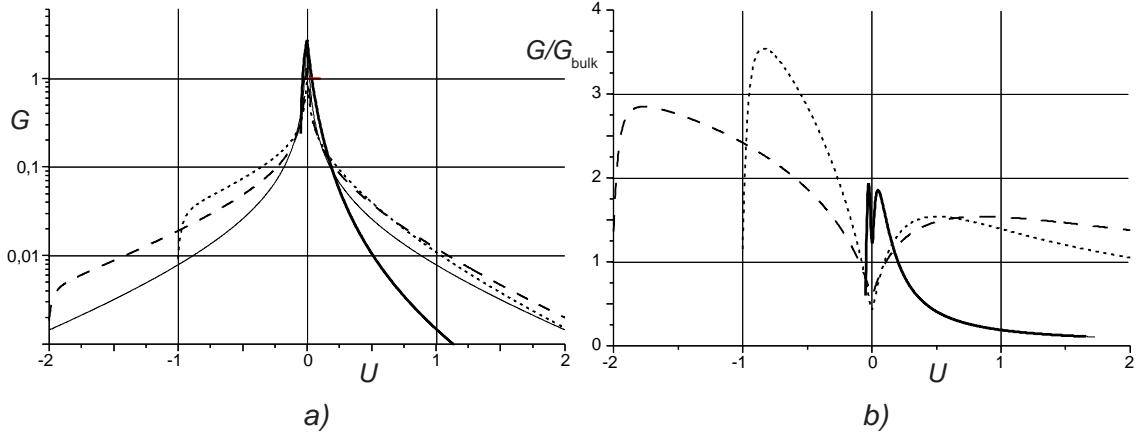, width=15cm} \caption{a) The correlation functions
$G(U,V,R)$ for $\xi=10^{-4},\; R=0.01$ and $V=0.05$ (dotted line), $V=1$ (dashed
line) and $V=2$ (dot-dashed line) as compared to the bulk correlation function for
similar $R$ (bold line). All the curves are normalized by the condition $\int
G(U,V,R) dU=1$; b) The ratios $G(U,V,R)/G_{\rm bulk}(U,R)$ for values of $\xi,\; R$
and $V$ similar to that in a).} \label{fig:7}
\end{figure}

Figure \ref{fig:7} shows the dependencies $G(U)$ for different values of $V$ as
compared with the bulk behavior. One sees easily how the presence of the wall
affects the correlation function. In particular, it is interesting to note the
increase of the polymer density in the vicinity of the wall, and also the smaller
increase of the density in the outer region due to polymer--wall hardcore repulsion.
The rapid decrease of the correlation function for $V=0.05$ with the growth of $U$
is due to the fact that in the system controlled by the fugacity $\epsilon$, the
mean size of clusters in the vicinity of the wall is much smaller than that in the
bulk (this is clearly outlined by the fact that $t(x) \rightarrow 1$ as $x
\rightarrow 0$).

\section{Discussion}

Summarizing, in this paper we presented a rigorous procedure to
describe the behavior of the large ideal trees near the
impenetrable (non-adsorbing) wall and obtained rather accurate
approximate expressions both for the partition function and
2-point correlation functions of the system.

For comparison of our results with those obtained
earlier \cite{debell,janssen} we need a better understanding of
the applicability of the original approximations of "large" and
"no" excluded volume.

For simplicity, let us fully neglect the attractive interactions
between the monomers. In this case the only volume interactions in
the system are those due to the excluded volume. To estimate how
much these interactions disturb the original conformation of the
non-interacting polymer one should (see \cite{gr_book}) calculate
the mean number of pairwise contacts of the monomers. This value
equals $n_2=\rho N$, where $\rho$, which is the mean density of
the polymer, can be calculated as $\rho=N
v/R_g^D\sim(v/a^3)N^{1/4}$, where $v$ is the excluded volume per
monomer. One can estimate that the theory presented above should
be valid in case when the number of pairwise contacts (and
therefore, as one can easily see, also the contacts of higher
order, i.e. triple, etc.) is negligible, or, in other words, if
$N\ll(a^3/v)^{4/5}$. On the contrary, if $N\gg(a^3/v)^{4/5}$, we
expect the results of \cite{janssen} to be valid. Note, that the
experimental value of the key parameter $Li=v/a^3$ (we refer to it
as the Lifshitz parameter) can vary widely. E.g., for the lattice
model $Li\sim1$ and, therefore, practically any tree on a lattice
cannot be considered as an ideal one. On the contrary, for many
real polymer systems $Li\ll 1$. Indeed, imagine, for example, a
tree constructed of star-like monomers, whose arms are long
polymer chains with associating groups at the end. For the
polymers of such architecture one can expect the Lifshitz
parameter $Li$ to be much less then unity. Thus, though in the
limit of $N\rightarrow\infty$ the volume interactions are always
important, there can be a rather wide range of polymer sizes $1\ll
N\ll Li^{-4/5}$ where the results obtained above are correct.

As it is mentioned in the Introduction, the statistics of randomly branched ideal
polymer chain near the impenetrable boundary was studied by supersymmetric methods
in the work \cite{janssen}, where the authors follow the general scheme of the
supersymmetric "dimensional reduction" for branched random walks formulated for the
first time in \cite{parisi} and exploited later in \cite{shapir,cardy}.
The authors of the paper \cite{janssen} have computed many
thermodynamic properties of branched polymer chains near the
repulsive and the adsorbing impenetrable surfaces in 3D. For our
purposes the most important are those of these results, which are
related to the "non-adsorbing" (i.e. repulsive) regime.

In \cite{janssen} the authors have got the following results below
the adsorbing transition point for the quantities of our interest:
\be \left\{\begin{array}{l}
\theta=\frac{5}{2} \medskip \\
G_N(z,z') = \text{erfc}(\zeta-\zeta')+\text{erfc}(\zeta+\zeta')-2\text{erfc}(\zeta)+
16\Gamma^2\exp\left[\Gamma(\Gamma-2\zeta-2\zeta') \right]
\end{array} \right. \label{eq:jan}
\ee
where
$$
\zeta=z\sqrt{N},\;\; \zeta'=z'\sqrt{N},\;\; \Gamma=E\sqrt{N}
$$
and $E$ is some constant independent of $N$.

It can be seen that the surface critical exponent
$\theta=\frac{5}{2}$ obtained in our work (eqs. (\ref{eq:28}),
(\ref{eq:29})) coincides with the one of \cite{janssen} (computed
also in \cite{debell}). We can thus assume that this critical
exponent seems to be independent of volume interactions, therefore
being genuine for all $N\gg 1$ independently of the Lifshitz
parameter $Li$.

Now, as far as the correlation functions is concerned, their
behaviour depends on the regime significantly. Indeed,
(\ref{eq:jan}) suggests the characteristic length scale of the
problem to be of order $r \sim \sqrt{N}$ (this result being in
compliance with the gyration radius for the randomly branched
polymers with excluded volume as obtained in
 \cite{parisi}). On the contrary, our approach results into $r
\sim N^{1/4}$, as the integral in (\ref{eq:47}) is mostly defined
by its lower limit (see also (\ref{eq:29}), where this
characteristic scale is clearly outlined), in full coincidence
with that obtained for the characteristic scale for the
non-interacting randomly branched polymer in the bulk (compare to
\cite{daou-joan,zimm-stock} and equations (\ref{eq:cor_sph3}),
(\ref{eq:gyr_bulk}) in this paper).

\begin{acknowledgments}

The authors are grateful to J.-F. Joanny for valuable stimulating discussions.
M.V.T. thanks the laboratory LPTMS (Universit\'e Paris Sud, Orsay) for the financial
support during the period in which the part of the work was done.

\end{acknowledgments}

\end{document}